# Low-energy Free-electron Rabi oscillation and its applications

Yiming Pan[1,2], Bin Zhang[3], Daniel Podolsky[2]

[1]School of Physical Science and Technology and Center for Transformative Science, ShanghaiTech University, Shanghai 200031, China

[2] Department of Physics, Technion, Haifa 3200003, Israel

[3] Department of Electrical Engineering Physical Electronics, Tel Aviv University, Ramat Aviv 6997801, Israel

**Abstract**

We propose free-electron Rabi oscillation by creating an isolated two-level system in a synthetic energy space induced by laser. The $\pi/2$-pulse and $\pi$-pulse in synthetic Rabi dynamics can function as "beam splitters" and "mirrors" for free-electron interferometry, allowing us to detect local electromagnetic fields and plasmonic excitations. When the coupling field is quantized, we can observe quantum and vacuum Rabi oscillations of the two-level electron, which can be used to investigate the quantum statistics of optical excitations and electron-photon entanglement. Recent advances in laser control of electron microscopes and spectroscopes makes the experimental detection of synthetic Rabi oscillations possible. However, observing the quantum Rabi oscillation of electrons remains challenging. Our work has the potential to advance various fundamentals and applications of resonant light-matter interactions between low-energy electrons and quatum light.

Atoms and molecules[1–6] have internal degrees of freedom that can be readily manipulated by light beams. In constrast, free electrons lack interior states or resonances for laser manipulation[7–9], as the electron's spin [10] is not significantly responsive to light in free space. Therefore, a critical question arises: is it possible to create an isolated two-level system of free electrons that is amenable to control by light?

In free space, when an electron absorbs or emits a photon, its center of mass recoils. The electron-photon interaction can lead to coherent wavefunction modulation and diffraction, such as the photon-induced near-field electron microscopy (PINEM)[11–13], and the transverse Kapitza-Dirac effect (KDE)[9,14]. PINEM is an ultrafast electron imaging technique that involves multiphoton emission and absorption[13]. Recent studies have rapidly advanced with the simplified picture of multi-level Rabi oscillation[15], such as free-electron quantum optics[16–20], free-electron qubits[21,22], free-electron-bounded-electron intractions[23,24] and PINEM synthetic dimensions[25–27]. Recent progress suggests that we are nearly on the same level of controlling and steering swift electrons as we in manipulating cold atoms with light[7,8]. Given the similar growth path, we can draw many concepts and applications from atom optics[1,5] and cavity QED[28,29] to study the expanding field of free electron optics[9,18,30,31]. However, at the most fundamental level, a spatially nonseparable isolated two-level electron system that is controllable by light is lacking.

Here, we construct a two-level system using laser-induced low-energy electrons. As the electron slows down (with a kinetic energy around 100 eV)[32,33], it only hops between two neighboring photon sidebands due to the quadratic dispersion of slow electrons, a phenomenon known as the Bragg diffraction. This is widely seen in many systems, such as atom optics[1], and acousto-optical modulators[34]. In contrast, the generation of infinite sidebands in the regular PINEM belongs to the Raman-Nath regime[32].

We demonstrated synthetic Rabi oscillation when the two-level electron interacts with a classical optical field, and predicted quantum and vacuum Rabi oscillations when the field is further quantized. These Rabi oscillations have potential uses, including building blocks for a novel free-electron interferometry and for sensing quantum statistics of optical fields and plasmonics[35]. The detection of vacuum Rabi oscillation may be used to explore many fundamental issues, such as electron-photon entanglement[17,31,36,37]. Our findings pave the way for future studies into these intriguing free-electron Rabi oscillations in low-energy electron quantum optics.

**Setup of low-energy electron Rabi oscillation** – The PINEM process of electrons interacting with an electromagnetic field can be described by a time-dependent Schrödinger equation: $i\hbar \frac{\partial}{\partial t}\psi(z,t) = (H_0 + H_I)\psi(z,t)$. The electron wavefunction can be expanded using the Floquet-Bloch basis as: $\psi(z,t) = \sum_n c_n(t) e^{-i(\omega_n t - k_n z)}$, with $\omega_n = \omega_0 + n\omega_L$ and $k_n = k_0 + nq_z$, where n is an integer or a half-integer. Here, $\omega_0 = \frac{\varepsilon_0}{\hbar}, k_0 = \frac{p_0}{\hbar}$ correspond to the synchronized kinetic energy and momentum of the electrons, while $\omega_L$ and $q_z = 2\pi/\Lambda$ corresponds to the laser frequency and the reciprocal wavevector of a grating with a period $\Lambda$ along the z axis. Under the synchronization (the group velocity of the electorn equals the phase velocity of the light field, $v_0 = \omega_L/q_z$) and nonrecoil approximations, we obtain the effectively coupled-mode equations:

$$i\frac{\partial}{\partial t} c_n = n^2 \epsilon\, c_n + \kappa c_{n+1} + \kappa^* c_{n-1}, \tag{1}$$

where the on-site potential $\epsilon = \frac{\hbar q_z^2}{2m_e}$, is determined from the quadratic dispersion curvature[32], and the hopping amplitude $\kappa = \frac{ek_0 E_z}{2m_e \omega_L}$, depends on the field strength $E_z$. When the spectral confinement is strong ($\epsilon \rangle \kappa$), we can truncate the infinite PINEM lattice into two sidebands. The explicit modeling is provided in the Method section. Therefore, we obtain an effective isolated two-level system:

$$\begin{aligned} i\frac{\partial}{\partial t} c_{\frac{1}{2}} &= \frac{\epsilon}{4} c_{\frac{1}{2}} + \kappa c_{-\frac{1}{2}} \\ i\frac{\partial}{\partial t} c_{-\frac{1}{2}} &= \frac{\epsilon}{4} c_{-\frac{1}{2}} + \kappa c_{\frac{1}{2}} \end{aligned} \tag{2}$$

where the two-level electron state $|\Psi(t)\rangle = c_{1/2}(t)\left|\frac{1}{2}\right\rangle + c_{-1/2}(t)\left|-\frac{1}{2}\right\rangle$, contains only two PINEM sidebands and all high-order terms are disregarded. We can solve the state $|\Psi(t)\rangle = e^{-\frac{i\epsilon t}{4}}\cos(\kappa t)\left|\frac{1}{2}\right\rangle - ie^{-\frac{i\epsilon t}{4}}\sin(\kappa t)\left|-\frac{1}{2}\right\rangle$. If the upper sideband is initially excited $|\Psi(0)\rangle = \left|\frac{1}{2}\right\rangle$, then after time t, the probability of finding the electron still there is given by:

$$P_{\frac{1}{2}}(t) = \frac{1}{2}(1 + \cos 2\kappa t). \tag{3}$$

This results in the occurrence of Rabi oscillations in a synthetic PINEM space. The duration of a Rabi cycle is $T_R = \pi/\kappa = \frac{2\pi\hbar\omega_L}{eE_zv_0}$, which represents the time required for an electron with velocity $v_0$ to absorb (or emit) a photon with energy $\hbar\omega_L$ in the presence of the field with the strength $E_z$.

Figure 2a shows the results obtained from direct numerical simulation of the time-dependent Schrödinger equation (TDSE). We see that the electron mainly oscillates between the two sidebands $\left|\pm\frac{1}{2}\right\rangle$, although the high-order terms $\left|\pm\frac{3}{2}\right\rangle$ also contribute to the coupling process to some extent. More information on the TDSE simulation is provided in the supplementary materials (SM) file.

The region of free electron and light interaction is determined by the ratio $Q = \frac{\epsilon}{2\kappa}$. Keeping the interaction in the Bragg regime requires Q to be large enough, and we need to carefully choose the experimental parameters. Indeed, we can express $Q$:

$$Q = \frac{\epsilon}{2|\kappa|} = \left(\frac{\hbar^2}{2m_e ec^3}\right)\frac{\omega_L^3}{\beta^3 E_z}. \tag{4}$$

Here we use the synchronization $q_z = \omega_L/\beta c$ for the 0-th sideband ($n = 0$), where except for the constants $\hbar, e, m_e, c$, we can tune Q by adjusting the frequency $\omega_L$, the field strength $E_z$, and the relative velocity $\beta$. As shown in Fig. S1 in the SM file, Q increases when the electron slows down. Fig. 2a illustrates the Rabi dynamics of a well isolated two-level electron. We chose $\beta = 0.02$, $\omega_L = 9.42$ PHz (the wavelength $\lambda_L = 200 nm$), and $E_z = 5 \times 10^6\ V/m$, resulting in Q = 29.3. Consequently, the Rabi cycle is $T_R = 1.29\ ps$, which corresponds to the interaction length of $7.74\ \mu m$.

**Free-electron interferometry -** A direct application of synthetic Rabi oscillation is to construct free-electron interferometry, which requires a "beam splitter" and a "mirror (or deflector)"[38–41]. Two Rabi cycles, namely, "π/2"- and "π"- pulses, are considered. Fig. 2b shows that a π/2-pulse creates an equally weighted superposition of $\left|+\frac{1}{2}\right\rangle$ and $\left|-\frac{1}{2}\right\rangle$, while a π-pulse flips them. Thus, a free-electron Mach-Zehnder interferometer can be constructed to measure the relative phase of a flying two-level electron gathered through local electromagnetic forces or gravity.

Figure 2c depicts an interferometer with a "π/2-π-π/2" sequence. Initially, an electron is prepared at $\left|+\frac{1}{2}\right\rangle$. Then, a π/2-pulse is used to split the electron into two sidebands. The electron travels through a region with a local electromagnetic potential, where it records a relative phase ($\Delta\phi$) due to the different momenta of the two sidebands. Next, a π-pulse deflects the electron, causing the populations of two sidebands to interchange. After free propagation over the same distance L, a π/2-pulse is used again to recombine the sidebands, converting the phase information back into state populations that can be read directly via an electron spectrometer[15].

The visibility of state population in the preceding "π/2-π-π/2" interferometry is given by $V = \left(P_{\frac{1}{2}} - P_{-\frac{1}{2}}\right) / \left(P_{\frac{1}{2}} + P_{-\frac{1}{2}}\right) = \cos\Delta\phi$. An analysis shows that the phase difference $\Delta\phi$ is affected by the local electromagnetic field and the acceleration of gravity. However, the contribution of gravity to $\Delta\phi$ is usually negligible, even for a slow electron flying with the speed of $c/100$ at energy of 100 eV, which would require a propagation distance of kilometers to accumulate a significant phase shift from gravity's acceleration. In contrast, because the two-level electron is negatively charged ($e^-$), it can accumulate a considerable phase from the coupling with the local and temporary fields in nanscales. Therefore, it is an promising sensor for detecting optical excitations, which can be used for developing ultrafast electron quantum sensors and interferometry with femtosecond temporal resolution and atomic spatial precision. Notably, Bragg diffraction of electron beam with transverse KDE modulation using pondermotive potential of light is also possible[7–9]. However, the KDE-induced sidebands are spatially separated in the transverse direction, while our longitudinal low-energy electron is not.

**Quantum collapse and revival -** When the optical field on the grating is quantized, we can obtain a two-level free-electron quantum optics Hamiltonian given by

$$H = \frac{\hbar v_0 q}{2}\left(c^\dagger_{\frac{1}{2}} c_{\frac{1}{2}} - c^\dagger_{-\frac{1}{2}} c_{-\frac{1}{2}}\right) + \hbar\omega a^\dagger_q a_q + g c^\dagger_{\frac{1}{2}} c_{-\frac{1}{2}} a_q + g^* c^\dagger_{-\frac{1}{2}} c_{\frac{1}{2}} a^\dagger_q \tag{5}$$

where the quantized coupling strength $g = \frac{ek_0\tilde{E}_z}{2\gamma m\omega_q}$ depends on the specific interaction configuration. Note that the Bragg condition $Q \gg 1$ is immediately satisfied for inputs of weak quantum light with a small averaged photon number $N_{ph}$ because $g\sqrt{N_{ph}} \ll \epsilon$. The second quantization procedure is provided in the SM file.

The Hamiltonian (Eq. 5) can be diagonalized in the same way as the celebrated Jaynes-Cummings model[33,42], since the equations are closed under the subspace $\left\{\left|\frac{1}{2}, v\right\rangle, \left|-\frac{1}{2}, v+1\right\rangle\right\}$. We define a general initial photon state $|\psi_{ph}\rangle = \sum_v c_v |v\rangle$, with $v$ being the Fock state index, and photon statistics given by $P_v = |c_v|^2$. Therefore, we obtain the quantum Rabi dynamics of the combined electron-photon state:

$$|\Psi(t)\rangle = \sum_v e^{-i\left(v+\frac{1}{2}\right)\omega t} c_{v+1}\left(\cos(g\sqrt{v+1}t)\left|\frac{1}{2}, v\right\rangle - \sin(g\sqrt{v+1}t)\left|-\frac{1}{2}, v+1\right\rangle\right), \quad (6)$$

from which the state population $P_{\frac{1}{2}}(t)$:

$$P_{\frac{1}{2}}(t) = \sum_v |c_{v+1}|^2 \cos^2(g\sqrt{v+1}t) = \sum_v P_{v+1} \cos^2(g\sqrt{v+1}t). \quad (7)$$

Figure 3 shows the Fourier transform of the inversion $I(t) = P_{\frac{1}{2}}(t) - P_{-\frac{1}{2}}(t)$ for different Rabi frequency order n, corresponding to three different photon statistics: thermal, coherent, and squeezed coherent. The numerical results, shown as dotted lines, demonstrate that the photon statistics of the light inputs can be extracted. In particular, the inset of Fig. 3 shows that for the small coherent state of light the pulse envelope of $I(t)$ dies and reappears, which is a phenomenon known as "collapse and revivals"[43] [4,44,45]. Three features of the collapse and revival are worth noting. First, the collapse time is determined by the spontaneous emission given by $t_c = \pi/g$ (which is the cycle of vacuum Rabi oscillation, we will discuss in the following section). Second, the revival period, given by $\tau_r = 4\pi\langle N_{ph}\rangle/g$, is linearly dependent on the input photon number. Third, this quantum Rabi oscillation differs from the synthetic oscillation in that the pulse never decays. However, in the semiclassical limit where $g \to 0$, $\langle N_{ph} \rangle \to \infty$, and $g\sqrt{N_{ph}} \to \kappa$, the quantum Rabi oscillation becomes the synthetic one (Eq. 3).

It should be noted that this approach differs from recent work on imprinting photon statistics on PINEM patterns[18,46], and while trapped ions and cavity QED[44,47,48] have been extensively studied, the use of free electron Rabi oscillations to detect the quantum nature of optical fields has not been previously reported.

**Spontaneous radiation-reaction cycling -** The vacuum Rabi oscillation is the opposite limit to the semiclassical limit [4,49,50]. When there is no laser input, we can obtain $P_{\frac{1}{2}}(t) = \cos^2(gt)$, and the oscillation cycle is $T_{VRO} = \pi/g$. Figure 4 illustrates how the two states $\left|\frac{1}{2}, 0\right\rangle, \left|-\frac{1}{2}, 1\right\rangle$ alternate. An electron that is initially in the upper sideband will transition to a lower sideband and emit a photon to the vacuum, creating a combined state $\left|-\frac{1}{2}, 1\right\rangle$. Later, the lower band electron can reabsorb the photon and bounce back to the upper one while the field also returns to its initial vacuum state. Without considering decoherence (e.g., damping, relaxation, and dephasing), we can assert the combined state cycles between $\left|\frac{1}{2}, 0\right\rangle$, and $\left|-\frac{1}{2}, 1\right\rangle$. This spontaneous electron-photon cycling is a pure quantum effect, which may imply a quantum analogy for the Abraham-Lorentz force (a self-force that occurs when an accelerating charge emits an electromagnetic field that acts back on the charge)[51–57].

Interestingly, the entanglement between electrons and photons prevents the detection individual photons. During spontaneous cycling, post-selecting the electron at $\left|\frac{1}{2}\right\rangle$ inhibits photon emission, while a "free" photon can be captured by a photodetector if the electron is selected at $\left|-\frac{1}{2}\right\rangle$. Feist et al.[31], demonstrated the coincidence of such an entangled electron-photon state through a combined measurement. Even with a non-fluorescence screen, post-selection would release a photon entangled with a two-level electron in a "metastable" state, supporting an interesting quantum interpretation of the unresolved Schwarz-Hora effect[58,59]. The free-electron vacuum Rabi oscillation shed new light on many quantum and classical electrodynamics concerns[36,60], despite potential controversy.

**Experimental considerations** – There are several challenges and opportunitiesin the experiments and fundamentals of confining the infinite PINEM synthetic space into a two-level system. The explicit condition $\Delta\omega_L < \kappa < \epsilon$ has to be scrutinized, which pertains to the frequency detuning ($\Delta\omega_L = \omega_L - v_0 q_z$), the coupling strength ($\kappa$), and the on-site potential ($\epsilon$). One aspect is that the applied laser field should be weak compared to the on-site potential ($|\kappa| < \epsilon$). However, a weak laser field increases the period of the Rabi oscillation, leading to more severe decoherence. Tuning down the laser frequency decreases $|\kappa|$ but also reduces the on-site potential, eventually violating the isolation of two sidebands. The other aspect is $\Delta\omega_L < |\kappa|$, is also challenging, corresponding to the phase-matching condition of low energy electron. However, periodic nanostructures and photonic crystals can be used to mediate the phase-matched interaction artificially [61–63], enabling customizable photonic modes with programmable frequency, wavevector, and polarization. For instance, assuming the optical harmonics $m = 5$ of a grating is synchronized, one Rabi cycle of the two-level electron corresponds to propagate 387 periods of the grating ($v_0 T_R/\Lambda = 387$), where the synchronized wavevector $q_z$ on the grating is given by $q_z = k_L \cos\theta + mq$, $\theta$ is the incident angle of the incident laser, and $q = \frac{2\pi}{\Lambda}$ is the reciprocal vector of the grating with period $\Lambda = m\beta\lambda_L = 20\ nm$ and m is the index of Floquet harmonics.

Second, box quantization in a vacuum is used to quantize the optical and plasmonic fields on a grating. Usually, simple quantization is insufficiently because these fields are temporally near-fields and coupled to a designed open interface of a nanomaterial. Establishing mesoscopic quantum electrodynamics may provide a suitable quantization and effective theoretical treatment on these local fields[35,64]. Our proposed free electron quantum Rabi oscillation (Fig. 3) may provide a novel technique to evaluate the quantum nature of these local fields.

Third, the vacuum Rabi oscillation (VRO) of an electron is inherently damped because it is caused by vacuum fluctuations of the optical fields even without a cavity. The field on a grating is different from an optical cavity, and thus decoherence and dephasing are inevitable. Therefore, measuring free-electron VRO is difficult and requires high-quality nanostructures to maintain quantum coherence. Notably, VRO can provide the tools for performing essential two-qubit logic operations[65] when two electrons are involved, offering the capability to develop a novel quantum information processing approach different from that in Ref: [21,22]. An analysis of free-electron universal logic operations will be conducted in our future studies.

To this end, we admit that our proposal of two-level electron is still a long way from being experimentally realized. Nevertheless, it is worthy noting the advantages between our approach and atom optics. As shown in Fig. 1c and 1d, the PINEM process for both slow and fast electrons is comparable to the two regimes in atom optics. Both electrons and atoms have the de Broglie wavelength in the range of 0.1-1 Å, roughly $10^4$ times smaller than an optical wavelength. The energy spacing of a two-level electron is a light quantum ($\hbar\omega$), while the internal state energy of atoms is less than 1 $\mu eV$ (in the rf regime). This makes the free-electron qubit detectable directly using an energy spectrometer, and more stable, eliminating the need for severe cooling and trapping techniques.

**Conclusion -** In brief, we created an isolated two-level system using the laser-shaped low energy free electrons, and proposed synthetic, quantum, and vacuum free-electron Rabi oscillations. The synthetic Rabi oscillation can be utilized to construct free-electron interferometry, which can detect local optical or plasmonic fields across femtosecond and atomic scales. The quantum Rabi oscillations of electrons can detect photon statistics of optical and plasmonic excitations. Additionally, the vacuum Rabi oscillation could be used to explore electron-photon entanglement and various contentious issues in electrodynamics. We expect that the two-level electrons can open up a promising route for exploring low energy free-electron optics and quantum simulations.

**Acknowledgments**

We thank Maor Elder and Nir Davidson for insightful discussions. This work was supported by the Israel Science Foundation (Grant No. 1803/18). The authors declare no competing financial interests.

Correspondence and requests for materials should be addressed to Y.P (yiming.pan@campus.technion.ac.il) and B. Z (binzhang@mail.tau.ac.il).

*Note 1: In addition, a detailed analysis of free-electron Rabi oscillations will appear soon.*

*Note 2: During the submission of our work, we find a similar work has posted on Arxiv by Aviv Karnieli, and Shanhui Fan, arXiv:2302.01575.*

**Methods**

**Two-level electrons** - A relativistically-modified Schrödinger equation of electrons may represent the PINEM process in the presence of the electromagnetic field,

$$i\hbar \frac{\partial}{\partial t}\psi(z,t) = (H_0 + H_I)\psi(z,t).$$

The free-electron Hamiltonian for one-dimensional relativistic dynamics is $H_0 = \varepsilon_0 + v_0(p - p_0) + \frac{(p-p_0)^2}{2\gamma^3 m}$, derived by expanding the Dirac equation with the initial momentum $p_0 = \gamma m v_0$ when the spin index is ignored. We choose the synchronized electron's kinetic energy $\varepsilon_0 = 100\ eV$, corresponding to the electron velocity $v_0 = \beta c$ with the relative speed $\beta = 0.02$ and the Lorentz factor $\gamma \approx 1$.. The near-field interaction is $H_I = -\frac{e}{2m}(A \cdot p + p \cdot A)$, without gauging, and the transverse field components are ignored. We take a realistic nanograting suggested in Ref [66], whose longitudinal electric field is approximately given by $E(z,t) = E_z \cos(\omega_L t - q_z z + \phi_0)$, and correspondingly the vector potential is $A(z,t) = -\frac{E_z}{\omega_L}\sin(\omega_L t - q_z z + \phi_0)$, with electric field strength $E_z$, laser frequency $\omega_L$, wavevector $q_z = \frac{2\pi}{\Lambda}$ for a grating with a period $\Lambda$, and the phase delay $\phi_0$. We can solve the time-dependent Schrödinger equation numerically[27], see the details in the SM file.

The electron wavefunction can be expanded on the Floquet-Bloch basis[27]

$$\psi(z,t) = \sum_n c_n(t) e^{-i(\omega_n t - k_n z)}, \tag{M1}$$

with the frequency, $\omega_n = \omega_0 + n\omega_L$ and the wave vector $k_n = k_0 + nq_z$. The notations $\omega_0 = \frac{\varepsilon_0}{\hbar}, k_0 = \frac{p_0}{\hbar}$ are the energy and momentum of the 0-th order sideband electron, and $n$ is an integer (or a half-integer) reflecting the number of photons emitted or absorbed by the electron. We take the synchronization condition $\omega_L - v_0 q_z \simeq 0$ for n=0, and the nonrecoil approximation $k_{n+1} - \frac{q_z}{2} \approx k_{n-1} + \frac{q_z}{2} \approx k_0$ since $k_0 \gg q_z$. With these approximations, we can obtain the coupled-mode equation (Eq.1 in the main text)

$$i\frac{\partial}{\partial t} c_n = n^2 \epsilon\, c_n + \kappa c_{n+1} + \kappa^* c_{n-1}, \tag{M2}$$

where the on-site potential is $\epsilon = \frac{\hbar q_z^2}{2\gamma^3 m}$, steming from the electron dispersion curvature, and the hopping amplitude between sidebands is $\kappa = -\frac{e k_0 E_z}{2\gamma m \omega_L} e^{i(\phi_0 + \frac{\pi}{2})}$, which depends on the laser field strength $E_z$. For simplicity, we choose $\phi_0 = \frac{\pi}{2}$ and then obtain the real value of $\kappa = \frac{e k_0 E_z}{2\gamma m \omega_L}$.

**References**:

1. Cronin, A. D., Schmiedmayer, J. & Pritchard, D. E. Optics and interferometry with atoms and molecules. *Rev Mod Phys* **81**, 1051 (2009).
2. Hornberger, K., Gerlich, S., Haslinger, P., Nimmrichter, S. & Arndt, M. Colloquium: Quantum interference of clusters and molecules. *Rev Mod Phys* **84**, 157 (2012).
3. Wineland, D. J. *et al.* Experimental issues in coherent quantum-state manipulation of trapped atomic ions. *J Res Natl Inst Stand Technol* **103**, 259 (1998).
4. Raimond, J.-M., Brune, M. & Haroche, S. Manipulating quantum entanglement with atoms and photons in a cavity. *Rev Mod Phys* **73**, 565 (2001).
5. Abend, S. *et al.* Atom interferometry and its applications. *Volume 197: Foundations of Quantum Theory* 345–392 (2019).
6. Wineland, D. J. *et al.* Experimental issues in coherent quantum-state manipulation of trapped atomic ions. *J Res Natl Inst Stand Technol* **103**, 259 (1998).
7. Batelaan, H. Colloquium: Illuminating the Kapitza-Dirac effect with electron matter optics. *Rev Mod Phys* **79**, 929 (2007).
8. Jones, E., Becker, M., Luiten, J. & Batelaan, H. Laser control of electron matter waves. *Laser Photon Rev* **10**, 214–229 (2016).
9. Freimund, D. L., Aflatooni, K. & Batelaan, H. Observation of the Kapitza–Dirac effect. *Nature* **413**, 142–143 (2001).
10. Batelaan, H., Gay, T. J. & Schwendiman, J. J. Stern-Gerlach effect for electron beams. *Phys Rev Lett* **79**, 4517 (1997).
11. Barwick, B., Flannigan, D. J. & Zewail, A. H. Photon-induced near-field electron microscopy. *Nature* **462**, 902–906 (2009).
12. Park, S. T., Lin, M. & Zewail, A. H. Photon-induced near-field electron microscopy (PINEM): theoretical and experimental. *New J Phys* **12**, 123028 (2010).
13. García de Abajo, F. J., Asenjo-Garcia, A. & Kociak, M. Multiphoton Absorption and Emission by Interaction of Swift Electrons with Evanescent Light Fields. *Nano Lett* **10**, 1859–1863 (2010).
14. Kapitza, P. L. & Dirac, P. A. M. The reflection of electrons from standing light waves. in *Mathematical Proceedings of the Cambridge Philosophical Society* vol. 29 297–300 (Cambridge University Press, 1933).
15. Feist, A. *et al.* Quantum coherent optical phase modulation in an ultrafast transmission electron microscope. *Nature* **521**, 200–203 (2015).
16. Kfir, O. *et al.* Controlling free electrons with optical whispering-gallery modes. *Nature* **582**, 46–49 (2020).
17. Kfir, O. Entanglements of Electrons and Cavity Photons in the Strong-Coupling Regime. *Phys Rev Lett* **123**, 103602 (2019).
18. Raphael, D. *et al.* Imprinting the quantum statistics of photons on free electrons. *Science (1979)* **373**, eabj7128 (2022).
19. A, B. H. *et al.* Shaping quantum photonic states using free electrons. *Sci Adv* **7**, eabe4270 (2022).
20. Gorlach, A. *et al.* Ultrafast non-destructive measurement of the quantum state of light using free electrons. *arXiv preprint arXiv:2012.12069* (2020).

21. Reinhardt, O., Mechel, C., Lynch, M. & Kaminer, I. Free-electron qubits. *Ann Phys* **533**, 2000254 (2021).
22. Tsarev, M. V, Ryabov, A. & Baum, P. Free-electron qubits and maximum-contrast attosecond pulses via temporal Talbot revivals. *Phys Rev Res* **3**, 43033 (2021).
23. Gover, A. & Yariv, A. Free-Electron--Bound-Electron Resonant Interaction. *Phys Rev Lett* **124**, 64801 (2020).
24. Zhao, Z., Sun, X.-Q. & Fan, S. Quantum Entanglement and Modulation Enhancement of Free-Electron--Bound-Electron Interaction. *Phys Rev Lett* **126**, 233402 (2021).
25. Braiman, G., Reinhardt, O., Levi, O., Mechel, C. & Kaminer, I. The Synthetic Hilbert Space of Laser-Driven Free-Electrons. in *CLEO: QELS_Fundamental Science* FTh1N-6 (Optica Publishing Group, 2021).
26. Pick, A., Reinhardt, O., Plotnik, Y., Wong, L. J. & Kaminer, I. Bloch oscillations of a free electron in a strong field. in *2018 Conference on Lasers and Electro-Optics (CLEO)* 1–2 (2018).
27. Pan, Y., Zhang, B. & Podolsky, D. Synthetic dimensions using ultrafast free electrons. *arXiv preprint arXiv:2207.07010* (2022).
28. Walther, H., Varcoe, B. T. H., Englert, B.-G. & Becker, T. Cavity quantum electrodynamics. *Reports on Progress in Physics* **69**, 1325 (2006).
29. Haroch, S. Mesoscopic State Superpositions and Decoherence in Quantum Optics. in *Les Houches* vol. 79 55–159 (Elsevier, 2004).
30. Kiesel, H., Renz, A. & Hasselbach, F. Observation of Hanbury Brown–Twiss anticorrelations for free electrons. *Nature* **418**, 392–394 (2002).
31. Feist, A. *et al.* Cavity-mediated electron-photon pairs. *Science (1979)* **377**, 777–780 (2022).
32. Eldar, M., Pan, Y. & Krüger, M. Self-trapping of slow electrons in the energy domain. *arXiv preprint arXiv:2209.14850* (2022).
33. Karnieli, A. & Fan, S. Jaynes-Cummings interaction between low energy free-electrons and cavity photons. *arXiv preprint arXiv:2302.01575* (2023).
34. Duocastella, M., Surdo, S., Zunino, A., Diaspro, A. & Saggau, P. Acousto-optic systems for advanced microscopy. *Journal of Physics: Photonics* **3**, 012004 (2020).
35. García de Abajo, F. J. Optical excitations in electron microscopy. *Rev Mod Phys* **82**, 209–275 (2010).
36. Pan, Y. & Gover, A. Spontaneous and stimulated emissions of a preformed quantum free-electron wave function. *Phys Rev A (Coll Park)* **99**, 52107 (2019).
37. Di Giulio, V. & García de Abajo, F. J. Free-electron shaping using quantum light. *Optica* **7**, 1820–1830 (2020).
38. Kasevich, M. & Chu, S. Atomic interferometry using stimulated Raman transitions. *Phys Rev Lett* **67**, 181 (1991).
39. Keith, D. W., Ekstrom, C. R., Turchette, Q. A. & Pritchard, D. E. An interferometer for atoms. *Phys Rev Lett* **66**, 2693 (1991).
40. Carnal, O. & Mlynek, J. Young's double-slit experiment with atoms: A simple atom interferometer. *Phys Rev Lett* **66**, 2689 (1991).

41. Riehle, F., Kisters, T., Witte, A., Helmcke, J. & Bordé, C. J. Optical Ramsey spectroscopy in a rotating frame: Sagnac effect in a matter-wave interferometer. *Phys Rev Lett* **67**, 177 (1991).
42. Jaynes, E. T. & Cummings, F. W. Comparison of quantum and semiclassical radiation theories with application to the beam maser. *Proceedings of the IEEE* **51**, 89–109 (1963).
43. Eberly, J. H., Narozhny, N. B. & Sanchez-Mondragon, J. J. Periodic spontaneous collapse and revival in a simple quantum model. *Phys Rev Lett* **44**, 1323 (1980).
44. Rempe, G., Walther, H. & Klein, N. Observation of quantum collapse and revival in a one-atom maser. *Phys Rev Lett* **58**, 353 (1987).
45. Haroche, S. & Raimond, J. M. Radiative properties of Rydberg states in resonant cavities. in *Advances in atomic and molecular physics* vol. 20 347–411 (Elsevier, 1985).
46. Di Giulio, V., Kociak, M. & de Abajo, F. J. G. Probing quantum optical excitations with fast electrons. *Optica* **6**, 1524–1534 (2019).
47. Brune, M. *et al.* Quantum Rabi oscillation: A direct test of field quantization in a cavity. *Phys Rev Lett* **76**, 1800 (1996).
48. Meekhof, D. M., Monroe, C., King, B. E., Itano, W. M. & Wineland, D. J. Generation of nonclassical motional states of a trapped atom. *Phys Rev Lett* **76**, 1796 (1996).
49. Agarwal, G. S. Vacuum-field Rabi splittings in microwave absorption by Rydberg atoms in a cavity. *Phys Rev Lett* **53**, 1732 (1984).
50. Agarwal, G. S. Vacuum-field Rabi oscillations of atoms in a cavity. *JOSA B* **2**, 480–485 (1985).
51. Jackson, J. D. Classical electrodynamics. Preprint at (1999).
52. Milonni, P. W. *The quantum vacuum: an introduction to quantum electrodynamics*. (Academic press, 2013).
53. Cole, J. M. *et al.* Experimental evidence of radiation reaction in the collision of a high-intensity laser pulse with a laser-wakefield accelerated electron beam. *Phys Rev X* **8**, 011020 (2018).
54. Poder, K. *et al.* Experimental signatures of the quantum nature of radiation reaction in the field of an ultraintense laser. *Phys Rev X* **8**, 031004 (2018).
55. Crisp, M. D. & Jaynes, E. T. Radiative effects in semiclassical theory. *Physical Review* **179**, 1253 (1969).
56. Stroud Jr, C. R. & Jaynes, E. T. Long-term solutions in semiclassical radiation theory. *Phys Rev A (Coll Park)* **1**, 106 (1970).
57. Ackerhalt, J. R., Knight, P. L. & Eberly, J. H. Radiation reaction and radiative frequency shifts. *Phys Rev Lett* **30**, 456 (1973).
58. Schwarz, H. & Hora, H. Modulation of an electron wave by a light wave. *Appl Phys Lett* **15**, 349–351 (1969).
59. Morokov, Y. N. Schwarz-Hora effect: present-day situation. in *ICONO'98: Fundamental Aspects of Laser-Matter Interaction and New Nonlinear Optical Materials and Physics of Low-Dimensional Structures* vol. 3734 34–40 (SPIE, 1999).
60. Pan, Y. & Gover, A. Beyond Fermi's golden rule in free-electron quantum electrodynamics: acceleration/radiation correspondence. *New J Phys* **23**, 063070 (2021).
61. Joannopoulos, J. D., Villeneuve, P. R. & Fan, S. Photonic crystals. *Solid State Commun* **102**, 165–173 (1997).

62. Lopez, C. Materials aspects of photonic crystals. *Advanced Materials* **15**, 1679–1704 (2003).
63. Lourtioz, J.-M. *et al.* Photonic crystals. *Towards Nanoscale Photonic Devices* (2005).
64. Rivera, N. & Kaminer, I. Light–matter interactions with photonic quasiparticles. *Nature Reviews Physics* **2**, 538–561 (2020).
65. Cirac, J. I. & Zoller, P. Quantum computations with cold trapped ions. *Phys Rev Lett* **74**, 4091 (1995).
66. Breuer, J., McNeur, J. & Hommelhoff, P. Dielectric laser acceleration of electrons in the vicinity of single and double grating structures—theory and simulations. *Journal of Physics B: Atomic, Molecular and Optical Physics* **47**, 234004 (2014).
67. Dickerson, S. M., Hogan, J. M., Sugarbaker, A., Johnson, D. M. S. & Kasevich, M. A. Multiaxis inertial sensing with long-time point source atom interferometry. *Phys Rev Lett* **111**, 083001 (2013).
68. Lamporesi, G., Bertoldi, A., Cacciapuoti, L., Prevedelli, M. & Tino, G. M. Determination of the Newtonian gravitational constant using atom interferometry. *Phys Rev Lett* **100**, 050801 (2008).
69. Kasevich, M. & Chu, S. Measurement of the gravitational acceleration of an atom with a light-pulse atom interferometer. *Applied Physics B* **54**, 321–332 (1992).
70. Kleinert, S., Kajari, E., Roura, A. & Schleich, W. P. Representation-free description of light-pulse atom interferometry including non-inertial effects. *Phys Rep* **605**, 1–50 (2015).
71. Kitching, J., Knappe, S. & Donley, E. A. Atomic sensors–a review. *IEEE Sens J* **11**, 1749–1758 (2011).
72. Fancher, C. T., Scherer, D. R., John, M. C. S. & Marlow, B. L. S. Rydberg atom electric field sensors for communications and sensing. *IEEE Transactions on Quantum Engineering* **2**, 1–13 (2021).
73. Fan, H. *et al.* Atom based RF electric field sensing. *Journal of Physics B: Atomic, Molecular and Optical Physics* **48**, 202001 (2015).

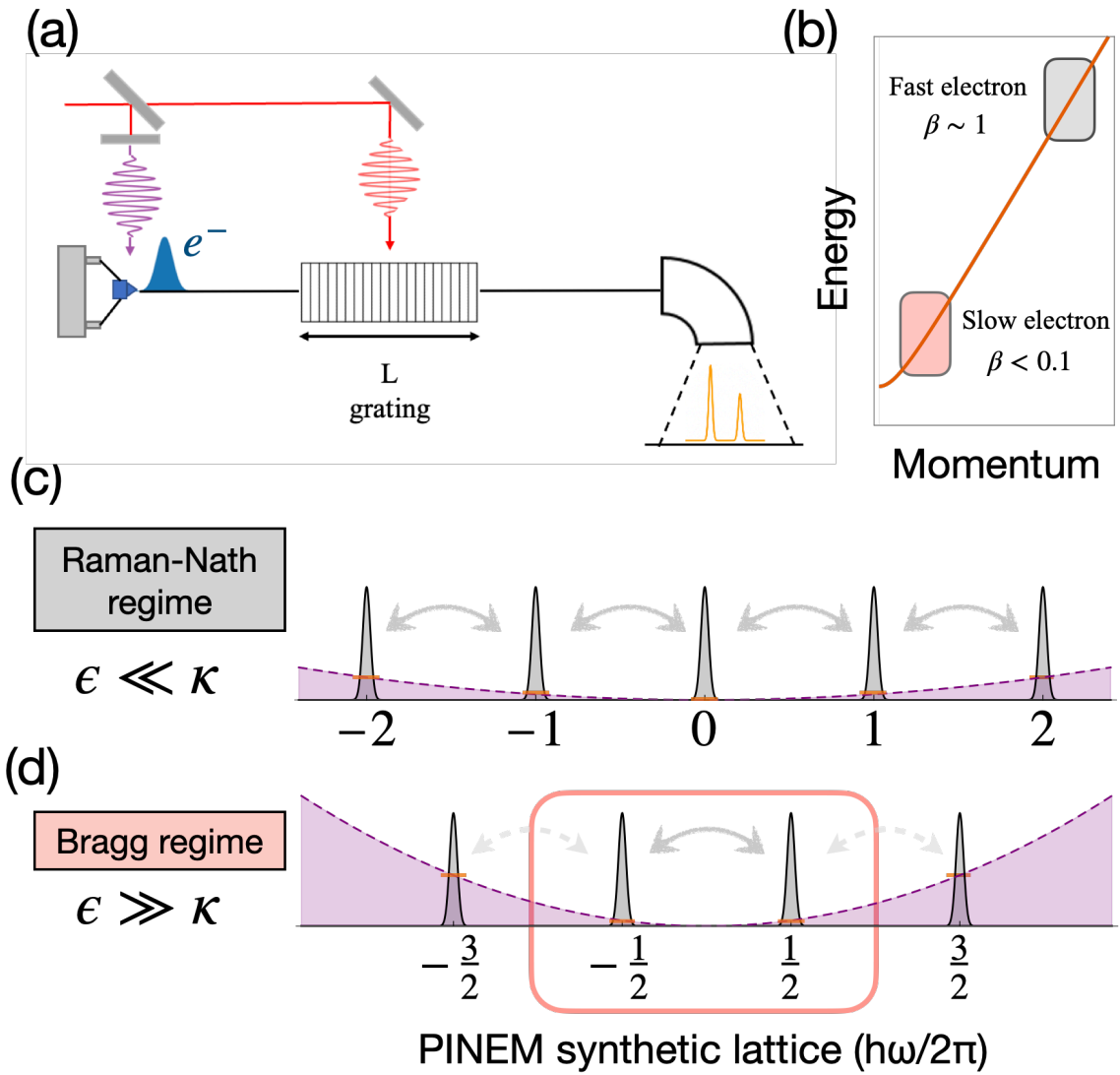

**Fig.1: Setup of slow electron Rabi oscillation in the synthetic energy lattice.** (a) The schematic diagram of the experimental setup includes the photoelectron gun, laser-illuminated grating, and high-resolution electron energy spectroscopy (EELS). (b) The energy-momentum dispersion with two regimes that are specifically marked. The slow electron regime is characterized by the relative velocity $\beta < 0.1$, while the fast electron is characterized by $\beta$ close to one. (c) The fast electron regime corresponds to the Raman-Nath approximation ($Q < 1$) in the PINEM synthetic lattice, and the slow electron regime corresponds to the Bragg diffraction regime ($Q > 1$), where an isolated two-level system is effectively formed.

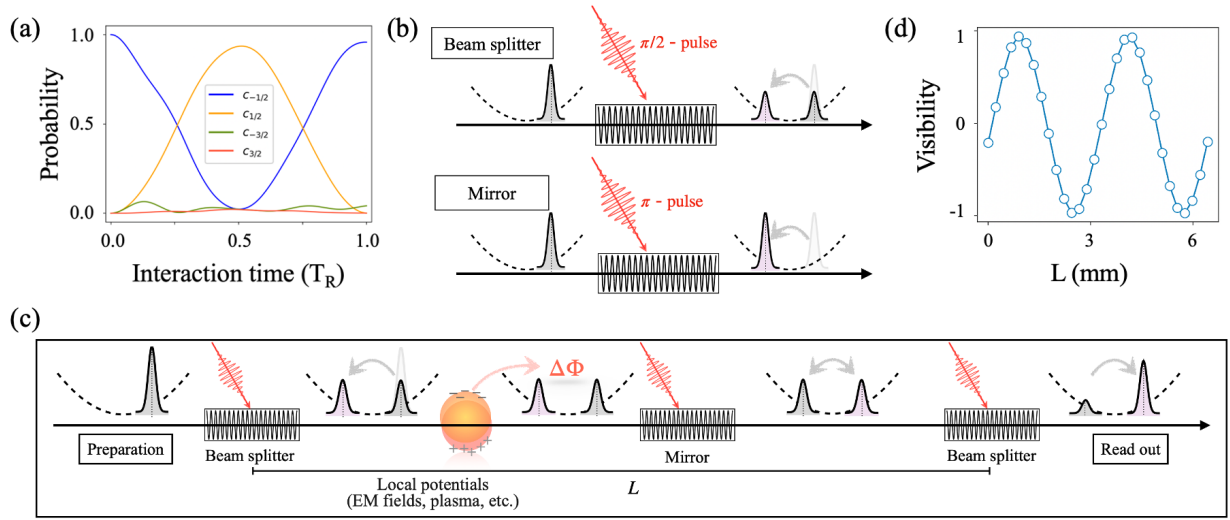

**Fig. 2: Synthetic Rabi oscillation and free electron optics.** (a) The Rabi oscillation between the two sidebands n=½ and n=–½. Here, we present direct solutions of TDSE when $Q = 29.3$, with the participation of high-order terms. (b) A beam splitter and a mirror for free electron wavefunction are constructed based on the $\frac{\pi}{2}$-pulse and $\pi$-pulse in the Rabi dynamics. (c) A free electron optics with a sequence of "π/2-π-π/2" is proposed to detect local potentials (local electromagnetic potential, plasmons, gravity, etc.) in analogy with atom optics. (d) The proposed interference fringes based on the local potentials (e.g., EM fields, plasma) are illustrated.

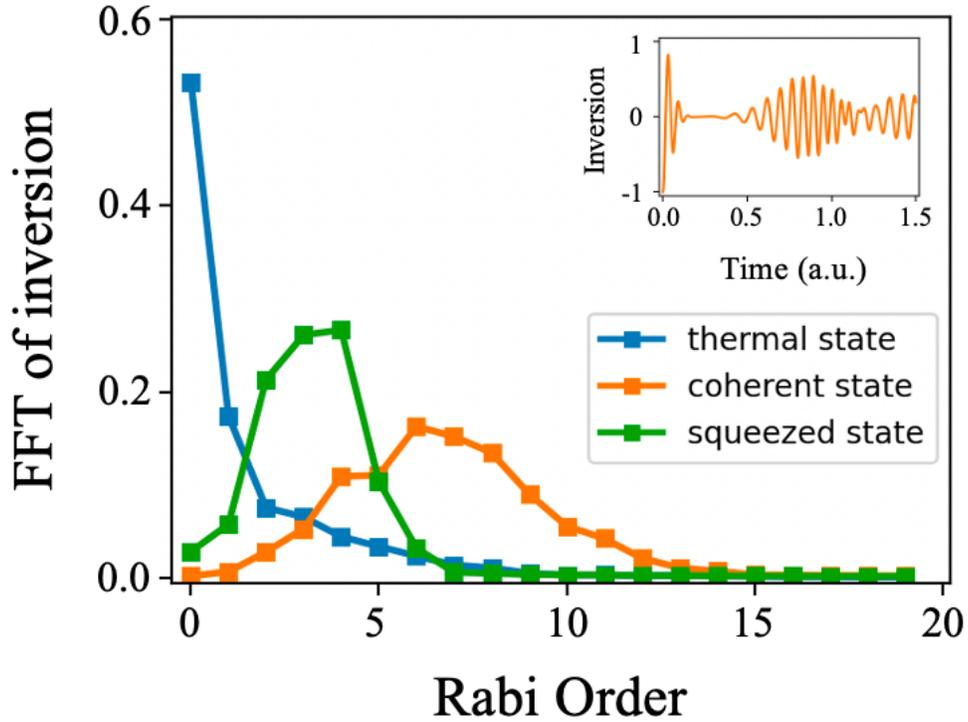

**Fig. 3: Direct measurement of photon statistics from the quantized Rabi oscillations.** The Fourier components of the inversion ($I(t) = P_{1/2} - P_{-1/2}$) as a function of the Rabi frequency order correspond to the input photon-number distributions of the thermal state (blue, with average photon number $\bar{n} = 2$), coherent state (orange, the amplitude $\alpha = 2.6$), and squeezed coherent state (green, $\alpha = 2.6$ and squeezing parameter $\xi = 0.4$), respectively. The Rabi dynamics of the free electron when coupling to the coherent state are shown in the inset, exhibiting the collapse and revival of quantum Rabi oscillation.

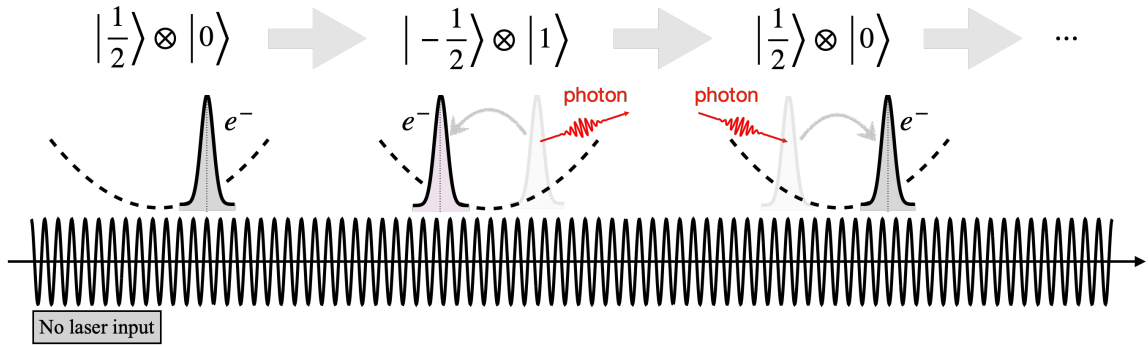

**Fig. 4: Vacuum Rabi oscillation of a slow electron and its spontaneous radiation-reaction cycling.** An electron is prepared at state $|½⟩$, and the photon is at vacuum $|0⟩$. The electron jumps to a lower energy state $\left|-\frac{1}{2}\right⟩$ and emits a photon, creating a combined state $\left|-\frac{1}{2}, 1\right⟩$. Then, the electron can reabsorb the emitted photon and jump back to the initial state $|½, 0⟩$. This cycling of photon radiation-electron reaction repeats periodically and spontaneously.

# Supplementary Material:

# Low-energy Free-electron Rabi oscillation and its applications

Yiming Pan[1,2], Bin Zhang[3], Daniel Podolsky[2]

[1]School of Physical Science and Technology and Center for Transformative Science, ShanghaiTech University, Shanghai 200031, China
[2] Department of Physics, Technion, Haifa 3200003, Israel
[3] Department of Electrical Engineering Physical Electronics, Tel Aviv University, Ramat Aviv 6997801, Israel

| Slow electron at 100 eV | Fast electron at 200 keV |
|---|---|
| Bragg regime (Q > 1) | Raman-Nath regime (Q < 1) |
| Rabi oscillation (two-level systems) | Multi-level Rabi oscillation (infinite ladders) |
| light-electron modulation (grating-based) | light-electron modulation (Tip-based) |
| Difficult to phase-matching | Easy to phase-matching |
| 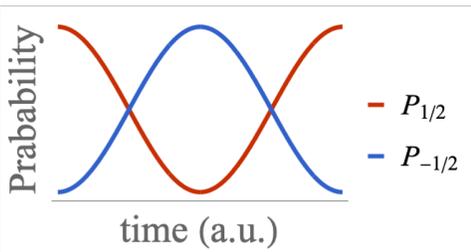 | 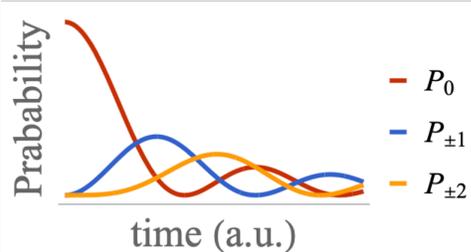 |

**Table 1:** Comparisons of the PINEM electron interaction with light in the Bragg and Raman-Nath regimes.

1. **The Bragg regime and Raman-Nath regime as a function for the relative velocity $\beta$ and field strength $E_z$**

According to the definition in the main text

$$Q = \frac{\epsilon}{2|\kappa|} = \left(\frac{\hbar^2}{2m_e ec^3}\right)\frac{\omega_L^3}{\beta^3\gamma^3 E_z} \tag{S1}$$

We show the density figure of $\ln Q$ as a function of the field strength $E_z$ and the electron velocity $\beta$. The electron interaction Bragg regime and Raman-Nath regime are divided by the dashed line $\ln Q = 0$. It is easier to realize the Bragg regime interaction (free-electron Rabi oscillation) when the electron velocity and the electric field strength are small.

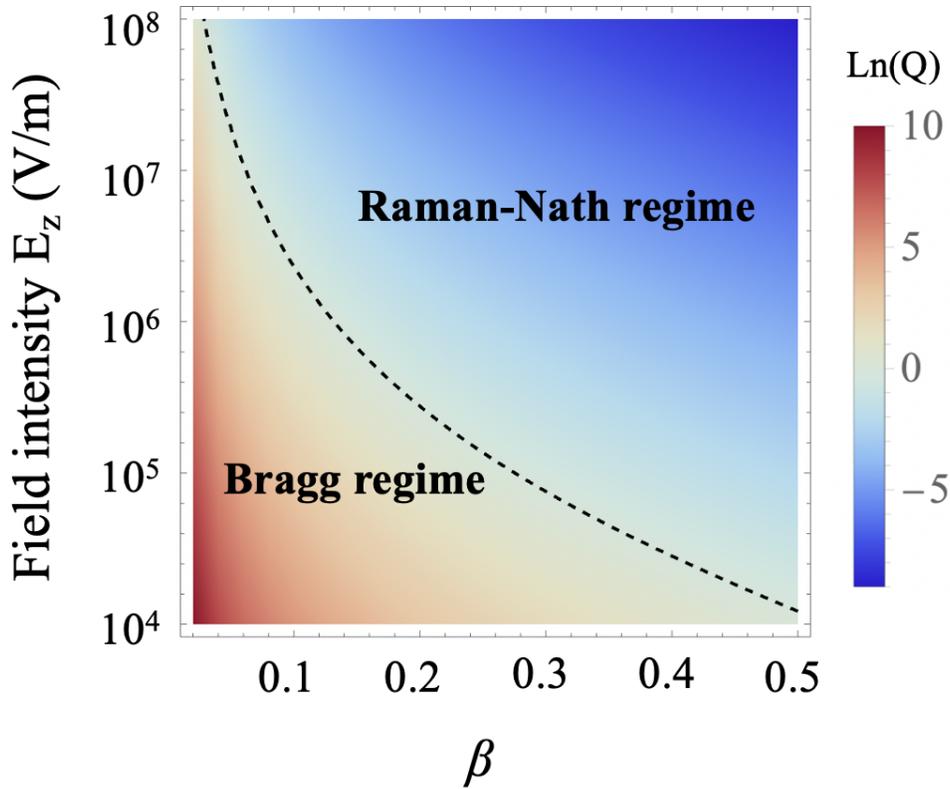

Fig. S1: The free electron interaction region is characterized by the parameter Q (color bar) regarding the relative velocity $\beta$ and the field amplitude $E_z$. $Q < 1$ corresponds to the Raman-Nath regime, whereas $Q > 1$ to the Bragg regime. For our concern, we choose the parameters $Q = 29.3$, and $|\kappa| = 0.0024$ fs$^{-1}$, $\epsilon = 0.142$ fs$^{-1}$) to obtain a two-level electron Rabi oscillation.

## 2. Comparison between free electron interferometry and atom interferometry

We send a two-level electron pass through a regime containing the electromagnetic field, static electric field, and gravity. Because the two sidebands relate to different electron momentum, they would accumulate different phases from the various interaction regimes. A simple theoretical analysis of the phase difference gives $\Delta\phi = -\frac{eEq}{2m}\left(\frac{L_2}{v_0}\right)^2 + \frac{qg}{2}\left(\frac{L_3}{v_0}\right)^2$, which contributes to the local electromagnetic vector potential $A$ (with the interaction length of $L_1$), the local electric field $E$ (different from the electric component of $A$, with the interaction length of $L_2$), and the acceleration of gravity $g$ (the interaction length of $L_3$), respectively.

Notably, we compare the potential sensitivity between our free-electron interferometry and atomic sensors in terms of the contributions from gravity, and electromagnetic fields, respectively. To begin, even if the flying qubit is a slow electron, it is still faster than atom-based qubits since the electron has a small mass. At 100 eV, the group velocity of a slow electron is $6 \times 10^6 m/s$; thus, this is much faster than the velocity of atoms, typically around $1\ mm/s$[67]. Even the atom's and electron's de Broglie wavelengths are near (both in the range of 0.1-1 Å, 10000 smaller than the light wavelength). Atom interferometry, in this sense, is a better approach to detect local gravitational accelerations[68], gravity gradients, and non-inertial effects[69,70] since the atoms are neutral and move slowly.

Different from atoms that are neutral particles, electrons are negatively charged, making local optical or plasmonic forces affect electrons considerably more strongly. Instead of sensing gravity's acceleration, electrons may be used to build quantum sensors for extremely weak and local electromagnetic fields such as near-fields, optical modes, and plasmonics. Besides, the current Rydberg atom antennas are more sensitive to electric fields ($\sim 10\ \mu V/cm$, the electric field of a single microwave photon confined in a wavelength volume) in the rf or microwave frequency range[71–73]. Meanwhile, the SQUID sensor, which uses a superconducting circuit, is the best sensor for detecting low magnetic fields.

So far, we have learned that free electron interferometry is far preferable for detecting local optical fields and plasmonic excitations at frequencies ranging from infrared to ultraviolet. The benefit of free electron interferometry is two-fold: it has a temporal resolution of the femtosecond scale (~fs) and a subatomic spatial resolution precision of the subatomic scale (~0.1nm). As a result, This makes it an ideal sensor for detecting local and transient electromagnetic excitations, which may compensate for the shortcomings and limits of other quantum sensors and interferometry.

## 3. Time-dependent Schrödinger equation direction simulation

In order to assess our tight-binding-approximated result of a two-level electron in the Bragg regime, we have to solve the time-dependent Schrödinger equation directly

$$i\hbar \frac{\partial}{\partial t} \psi(z,t) = \widehat{H}\psi(z,t) = (\widehat{H}_0 + \widehat{H}_I)\psi(z,t), \quad (S2)$$

where the kinetic Hamiltonian of free electron is $\widehat{H}_0 = E_0 + v_0(\hat{p} - p_0) + \frac{(\hat{p}-p_0)^2}{2\gamma^3 m_e}$. We choose the initial kinetic energy $\varepsilon_0 = 100\ eV$, corresponding to the electron velocity $v_0 = \beta c$ with the relative speed $\beta = 0.02$ and the Lorentz factor $\gamma = 1.001$. We take a realistic nanograting suggested in Ref[66], whose longitudinal vector potential is $A(z,t) = -\frac{E_z}{\omega_L}\sin(\omega_L t - q_z z + \phi_0)$, with electric field strength $E_z$, laser frequency $\omega_L$, wavevector $q_z = \frac{2\pi}{\Lambda}$ for a grating with a period $\Lambda$, and the phase $\phi_0$. Thus, the interaction Hamiltonian can be written as

$$\begin{aligned}\widehat{H}_I &= -\frac{e}{2\gamma m_e}[\hat{p}\cdot A(z,t) + A(z,t)\cdot \hat{p}]\\ &= -\frac{e}{\gamma m_e}A(z,t)\cdot\hat{p} - i\hbar k_z A_0 \cos(\omega_L t - k_z z + \phi_0) \\ &\simeq -\frac{eA_0}{\gamma m}\sin(\omega_L t - k_z z + \phi_0)\cdot\hat{p}\end{aligned} \quad (S3)$$

where we disregard the second term under the approximation $p_0 \gg \hbar k_z$ (this will be demonstrated in the following). We express the electron wavefunction as the slow-moving part with an initial phase:

$$\psi(z,t) = \chi(z,t)e^{-\frac{i(E_0 t - p_0 z)}{\hbar}} \quad (S4)$$

Thus, the LHS of Eq. (S2) is

$$i\hbar\frac{\partial}{\partial t}\psi(z,t) = e^{-\frac{i(E_0 t - p_0 z)t}{\hbar}}\left(E_0 + i\hbar\frac{\partial}{\partial t}\right)\chi(z,t) \tag{S5}$$

Meanwhile, the RHS is

$$\begin{aligned}&\left[e^{-\frac{i(E_0 t - p_0 z)t}{\hbar}}\left(E_0 + v_0((\hat{p}+p_0)-p_0) + \frac{((\hat{p}+p_0)-p_0)^2}{2\gamma^3 m_e}\right)\right.\\&\left.-e^{-\frac{i(E_0 t - p_0 z)t}{\hbar}}\frac{eA_0}{\gamma m_e}\sin(\omega_L t - k_z z + \phi_0)\cdot(\hat{p}+p_0)\right]\chi(z,t)\\&-\left[e^{-\frac{i(E_0 t - p_0 z)t}{\hbar}}i\hbar k_z\frac{eA_0}{2\gamma m_e}\cos(\omega_L t - k_z z + \phi_0)\right]\chi(z,t)\\&\simeq e^{-\frac{i(E_0 t - p_0 z)t}{\hbar}}\left(E_0 + v_0\hat{p} + \frac{\hat{p}^2}{2\gamma^3 m_e} - \frac{eA_0}{\gamma m_e}\sin(\omega_L t - k_z z + \phi_0)\cdot(\hat{p}+p_0)\right)\chi(z,t)\end{aligned} \tag{S6}$$

here we omit the third term since $\frac{i\hbar k_z}{2}\cos(\omega_L t - k_z z + \phi_0) \ll p_0 \sin(\omega_L t - k_z z + \phi_0)$ under $p_0 \gg \hbar k_z$. Thus, we simplify the TDSE as

$$\begin{aligned}i\hbar\frac{\partial}{\partial t}\chi(z,t) &= \left(\frac{\hat{p}^2}{2\gamma^3 m} + \left[v_0 - \frac{eA_0}{\gamma m}\sin(\omega_L t - k_z z + \phi_0)\right]\hat{p}\right)\chi(z,t)\\&-\frac{eA_0 p_0}{\gamma m}\sin(\omega_L t - k_z z + \phi_0)\chi(z,t)\end{aligned} \tag{S7}$$

Substitute the momentum operator $\hat{p} = -i\hbar\frac{\partial}{\partial z}$ and $p_0 = \gamma\beta mc$, $A_0 = \frac{E_0}{\omega_L}$

$$\begin{aligned}i\hbar\frac{\partial}{\partial t}\chi(z,t) &= \left(-i\hbar\left[v_0 - \frac{eE_0}{\gamma m\omega_L}\sin(\omega_L t - k_z z + \phi_0)\right]\frac{\partial}{\partial z}\right)\chi(z,t)\\&-\left[\frac{\hbar^2}{2\gamma^3 m}\frac{\partial^2}{\partial z^2} + \frac{eE_0\beta c}{\omega_L}\sin(\omega_L t - k_z z + \phi_0)\right]\chi(z,t)\end{aligned} \tag{S8}$$

Divided by $\hbar c$ into both sides

$$\begin{aligned}i\frac{\partial}{\partial\tau}\chi(z,\tau) &= -i\left[\beta - \frac{eE_0}{\gamma mc\omega_L}\sin(k_L\tau - k_z z + \phi_0)\right]\frac{\partial}{\partial z}\chi(z,\tau)\\&-\left[\frac{\hbar}{2\gamma^3 mc}\frac{\partial^2}{\partial z^2} + \frac{eE_0\beta}{\hbar\omega_L}\sin(k_L\tau - k_z z + \phi_0)\right]\chi(z,\tau)\end{aligned} \tag{S9}$$

where $\tau = ct$ and $k_L = \frac{\omega_L}{c}$. Notice that all the variables are calculated in the length unit of $\mu m$ and the time of $fs$. Assume that the electric field on the grating surface is $0.5 \times 10^7 \, V/m$ and the incident laser $\lambda_L = 0.2 \mu m$. Define

$$\alpha_1 = \frac{eE_0}{\gamma mc\omega_L} = 3.11 \times 10^{-7}, \quad \alpha_2 = \frac{\hbar}{2\gamma^3 mc} = 1.92 \times 10^{-7} \mu m, \quad \alpha_0 = \frac{eE_0 \beta}{\hbar \omega_L} = 0.016 \, \mu m^{-1}$$

As a result, the simplified equation is

$$i\frac{\partial}{\partial \tau}\chi(z,\tau) = -i[\beta - \alpha_1 \sin(k_L\tau - k_z z + \phi_0)]\frac{\partial}{\partial z}\chi(z,\tau)$$
$$- \left[\alpha_2 \frac{\partial^2}{\partial z^2} + \alpha_0 \sin(k_L\tau - k_z z + \phi_0)\right]\chi(z,\tau) \quad (S10)$$

Then we apply the discretization for the spatial parameter $z$ by dividing the spatial domain into $N_z$ parts, with the interval $\delta z = (z_{max} - z_{min})/N_z$

$$\frac{\partial^2}{\partial z^2}\chi(z,t) = \frac{1}{\delta z^2}[\chi(z+\delta z,t) + \chi(z-\delta z,t) - 2\chi(z,t)]$$
$$-i\beta \frac{\partial}{\partial z}\chi(z,t) = -\frac{i\beta}{2\delta z}[\chi(z+\delta z,t) - \chi(z-\delta z,t)]$$
$$i\alpha_1 \sin(k_L\tau - k_z z + \phi_0)\frac{\partial}{\partial z} = -\frac{\alpha_1}{\delta z}\cos(k_L\tau - k_z z + \phi_0)\chi(z,t) \quad (S11)$$
$$+ \frac{\alpha_1}{2\delta z}\left[e^{i(k_L\tau - k_z z + \phi_0)}\chi(z+\delta z,t) + e^{-i(k_L\tau - k_z z + \phi_0)}\chi(z-\delta z,t)\right]$$

Thus, the discretized Hamiltonian becomes

$$i\frac{\partial}{\partial \tau}\chi(z,\tau) = \left[-\frac{\alpha_2}{\delta z^2} - \frac{i\beta}{2\delta z} + \frac{\alpha_1}{2\delta z}e^{i(k_L\tau - k_z z + \phi_0)}\right]\chi(z+\delta z,t)$$
$$+ \left[-\frac{\alpha_2}{\delta z^2} + \frac{i\beta}{2\delta z} + \frac{\alpha_1}{2\delta z}e^{-i(k_L\tau - k_z z + \phi_0)}\right]\chi(z-\delta z,t) \quad (S12)$$
$$\left[\frac{2\alpha_2}{\delta z^2} - \frac{\alpha_1}{\delta z}\cos(k_L\tau - k_z z + \phi_0) - \alpha_0 \sin(k_L\tau - k_z z + \phi_0)\right]\chi(z,t)$$

In order to compute the time evolution of the slow-varying part $\chi(z,\tau)$ for a given initial state $\chi(z,\tau_0)$, we define the vector wavefunction with $N_z$ components

$$v(\tau) = (\chi(z_1,\tau), \chi(z_1,\tau), \ldots, \chi(z_1,\tau))^T \quad (S13)$$

Then, the differential equation (S12) can be simplified as

$$i\frac{\partial}{\partial \tau}v(\tau) = H(\tau)v(\tau) \tag{S14}$$

Applying the time discretization for the interval $\tau - \tau_0$ by equally dividing it into $N$ parts with the subinterval $\Delta\tau = (\tau - \tau_0)/N_t$ and we use an implicit Crank-Nicholson integrator to propagate the vector wavefunction from one time step to the next. The formal solution to Eq. (S14) can be expressed in terms of the time evolution operator

$$v(\tau + \Delta\tau) = U(\tau + \Delta\tau, \tau)v(\tau) \tag{S15}$$

where the time evolution operator can be expressed as

$$U(\tau + \Delta\tau, \tau) = (1 + i\,\Delta\tau H(\tau)/2)^{-1}(1 - i\,\Delta\tau H(\tau)/2) \tag{S16}$$

The solution of Eq. S13 can give us the dynamics of the PINEM electron with an initial condition. The final wavefunction could be expressed as

$$v(\tau) = \prod_{n=1}^{N_t} U(\tau_0 + n\Delta\tau, \tau_0 + (n-1)\Delta\tau)\, v(\tau_0) \tag{S17}$$

The result is presented in Fig. 2a in the main text. Here we show the comparison of the TDSE simulation and the TBA model (exact two-level system) in Fig. S1, which match well with each other.

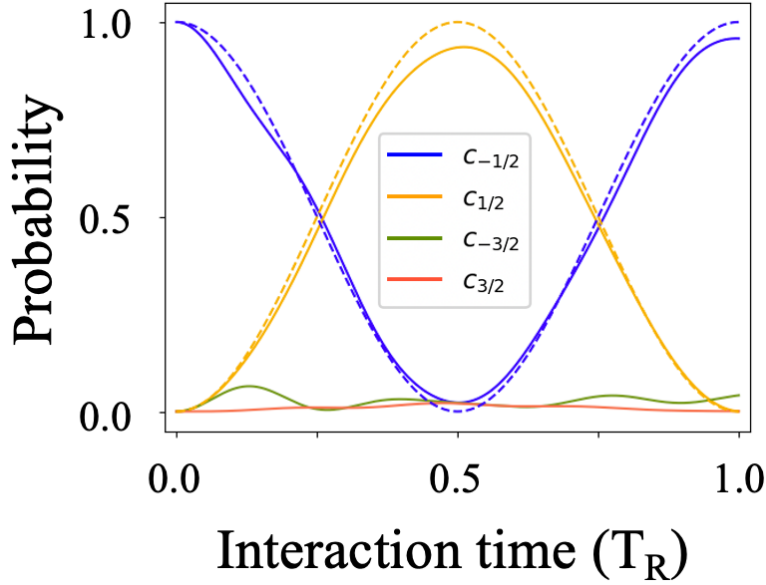

Fig. S2: TDSE simulation (solid line) of the time evolution of the PINEM electron. The electron wavefunction initially located at –½, exhibits Rabi oscillation between the two sidebands ½ and –½, which matches well with the exact Rabi oscillation of the TBA model. Part of the electron wavefunction is scattered into the 3/2 and –3/2, which is negligible.

## 4. Second-quantization procedure for quantum Rabi collapse and revival

Here we show the second quantization procedure of the electron-photon interaction, starting from the original one-dimensional relativistically-modified Schrödinger equation of the free electron in the presence of the electromagnetic field

$$i\hbar \frac{\partial}{\partial t}\psi(z,t) = (H_0 + H_p + H_I)\psi(z,t) \qquad (S18)$$

where $H_0 = \varepsilon_0 + v_0(p - p_0) + \frac{(p-p_0)^2}{2\gamma^3 m}$ is the kinetic energy of the free electron, $H_p$ is the total energy of the optically excited near-field on the grating, and $H_I$ is the interaction Hamiltonian between the free electron and the near-field. Different from Eq. (S1), here, the wavefunction represents the electron-photon state.

Introducing a general Hamiltonian of the excited near-field on a grating (we would specify this mode into a Floquet harmonics later.)

$$H_p = \frac{1}{2}\int dr \left[\frac{\varepsilon_0}{\eta^2} E(r,t)^2 + \mu_0 B(r,t)^2\right] \quad (S19)$$

With $E(r,t) = -\frac{\partial A(r,t)}{\partial t}, B(r,t) = \nabla \times A(r,t)$. And the interaction Hamiltonian has the form

$$H_I = -\frac{e}{\gamma m_e} A(z,t) \cdot p \quad (S20)$$

For the single-mode situation, the vector potential of the quantized field in 1D has the form

$$\hat{A}(z,t) = \frac{i\tilde{E}_z}{2\omega_q}\left[\hat{a}_q(t)e^{iqz} + \hat{a}_q^\dagger(t)e^{-iqz}\right] \quad (S21)$$

where $q_z$ is the wave vector of the optical excitation mode induced by the incident laser on the grating. For the Smith-Purcell type interaction, the phase matching condition is required in the electron propagation direction (z), which gives $q_z = \omega_L/v_0$. We can get the quantized Hamiltonian for photons through the quantized field $A(z,t)$

$$\hat{H}_p = \hbar\omega_q\left(\hat{a}_q^\dagger \hat{a}_q + \frac{1}{2}\right). \quad (S22)$$

Since the free electrons are moving through the periodic grating, the electron wavefunction can be expanded according to the Bloch mode

$$\psi(z,t) = \sum_n c_n(t)\, e^{ik_n z} \quad (S23)$$

where $k_n = (p_0 + n\hbar q_z)/\hbar$ is the electron momentum recoil due to the photon emission and absorption. The second quantization of the free electron can be applied by replacing the wavefunction $\psi(z,t)$ by field operator $\hat{\psi}(z,t)$, which satisfies the fermionic anticommutation relation $\{\hat{\psi}^\dagger(z,t), \hat{\psi}(z',t')\} = \delta(z-z')\delta(t-t')$.

Replacing the coefficient $c_n(t)$ with the annihilation operator $\hat{c}_n(t)$, which satisfies the anticommutation relation $\{\hat{c}_n^\dagger(t), \hat{c}_m(t')\} = \delta_{nm}\delta(t-t')$, we have

$$\hat{H}_e = \int dz\, \hat{\psi}^\dagger(z,t) H_e(p) \hat{\psi}(z,t)$$
$$= \int dz\, \hat{c}_n^\dagger \hat{c}_m \left[e^{-ik_n z} H_e\left(-i\hbar \frac{\partial}{\partial z}\right) e^{ik_m z}\right]$$
$$= \int dz\, \hat{c}_n^\dagger \hat{c}_m e^{i(k_m - k_n)z} \left[E_0 + v_0(\hbar k_m - p_0) + \frac{(\hbar k_m - p_0)^2}{2\gamma^3 m}\right]$$

Thus, the quantized kinetic Hamiltonian of the free electron can be written as

$$\hat{H}_e = \sum_n \varepsilon_n \hat{c}_n^\dagger \hat{c}_n \tag{S24}$$

where $\varepsilon_n = E_0 + v_0(\hbar k_n - p_0) + \frac{(\hbar k_n - p_0)^2}{2\gamma^3 m_e}$. Under the synchronism condition, we have $\varepsilon_n = E_0 + n\hbar v_0 q_z + \frac{n^2 \hbar^2 q_z^2}{2\gamma^3 m_e}$. With the quantization of both the free electron Eq. (S23) and the near field Eq. (S21) in the Schrodinger picture, we could derive the interaction Hamiltonian through Eq. (S20)

$$\hat{H}_I = -\frac{e}{\gamma m_e} \int dz\, \hat{\psi}^\dagger(z,t) \hat{A} \cdot \hat{p}\, \hat{\psi}(z,t)$$
$$= -\frac{e\tilde{E}_z}{2\gamma m_e \omega_q} \sum_{nm} \hat{c}_n^\dagger(t)\hat{c}_m(t) \int dz \left[\hat{a}_q(t)e^{iqz} + \hat{a}_q^\dagger(t)e^{-iqz}\right]\left(e^{-ik_n z}\, \hat{p}\, e^{ik_m z}\right)$$
$$= -\frac{e\tilde{E}_z \hbar}{2\gamma m_e \omega_q} \sum_{nm} \hat{c}_n^\dagger(t)\hat{c}_m(t)\, k_m \int dz \left[\hat{a}_q(t)e^{i(k_m - k_n + q_z)z} + \hat{a}_q^\dagger(t)e^{i(k_m - k_n - q_z)z}\right]$$
$$= -\frac{e\tilde{E}_z \hbar}{2\gamma m_e \omega_q} \sum_{nm} \hat{c}_n^\dagger(t)\hat{c}_m(t)\, k_m\left[\hat{a}_q(t)\delta(k_m - k_n + q_z) + \hat{a}_q^\dagger(t)\delta(k_m - k_n - q_z)\right]$$

According to the phase matching condition $\omega_L - v_0 q_z = 0$, the interaction Hamiltonian has the form

$$\hat{H}_I = -\frac{e\tilde{E}_z \hbar}{\gamma m_e \omega_q} \sum_n \left[k_{n-1} \hat{c}_n^\dagger \hat{c}_{n-1} \hat{a}_q(t) + k_{n+1} \hat{c}_n^\dagger \hat{c}_{n+1} \hat{a}_q^\dagger(t)\right] \tag{S25}$$

Since $q_z \ll k_m = k_0 + mq_z$, we assume that $k_{m-1} \simeq k_{m+1} \simeq k_0$. Thus, we can define the coupling constant

$$g = -\frac{e\tilde{E}_z}{2\gamma m_e \omega_q} k_0$$

The coupling strength corresponds to the Rabi frequency of vacuum Rabi oscillation. The field strength $\tilde{E}_z$ is the intensity of the electrical vacuum mode, which satisfies $E_z = \tilde{E}_z \langle a_q \rangle$ for the semiclassical field. In our simulation, we use $\tilde{E}_z = 10^5 \ V/m$ and the coupling constant $g = 0.2 \text{THz}$. The corresponding Rabi oscillation period $T_R = 16.2 \ ps$.

Then the total Hamiltonian can be written as

$$\hat{H} = \sum_{n=-\infty}^{\infty} \varepsilon_n \hat{c}_n^\dagger \hat{c}_n + \hbar\omega_q \left(\hat{a}_q^\dagger \hat{a}_q + \frac{1}{2}\right) + \hbar g \sum_n (\hat{c}_n^\dagger \hat{c}_{n-1} \hat{a}_q + \hat{c}_n^\dagger \hat{c}_{n+1} \hat{a}_q^\dagger) \quad (S26)$$

In the fully quantized case, when the parameters satisfy $\epsilon/2|g|$ "1, the system could also reach the Raman-Nath regime, which means only two sidebands are involved in the interaction (as shown in fig. S3). So, we could keep only two terms in Eq. (S26) ($m = \pm\frac{1}{2}$) and get the Hamiltonian in the two-level system representation

$$\begin{aligned} H &= \frac{\hbar v_0 q_z}{2}\left(c_{\frac{1}{2}}^\dagger c_{\frac{1}{2}} - c_{-\frac{1}{2}}^\dagger c_{-\frac{1}{2}}\right) + \hbar\omega a_q^\dagger a_q + \hbar\, g c_{\frac{1}{2}}^\dagger c_{-\frac{1}{2}} a_q + \hbar g^* c_{-\frac{1}{2}}^\dagger c_{\frac{1}{2}} a_q^\dagger \\ &= \hbar v_0 q_z \tau_z + \hbar\omega a_q^\dagger a_q + \hbar\left(g\tau_+ a_q + g^*\tau_- a_q^\dagger\right) \\ &= H_0 + H_I \end{aligned} \quad (S27)$$

where $H_0 = \hbar v_0 q_z \tau_z + \hbar\omega a_q^\dagger a_q$ is the free Hamiltonian of the electron and photon, $H_I = \hbar(g\tau_+ a_q + g^*\tau_- a_q^\dagger)$ is the interaction part. And

$$\tau_z = \frac{\sigma_z}{2} = \begin{pmatrix} 1/2 & 0 \\ 0 & -1/2 \end{pmatrix}, \tau_- = \begin{pmatrix} 0 & 0 \\ 1 & 0 \end{pmatrix}, \tau_+ = \begin{pmatrix} 0 & 1 \\ 0 & 0 \end{pmatrix}$$

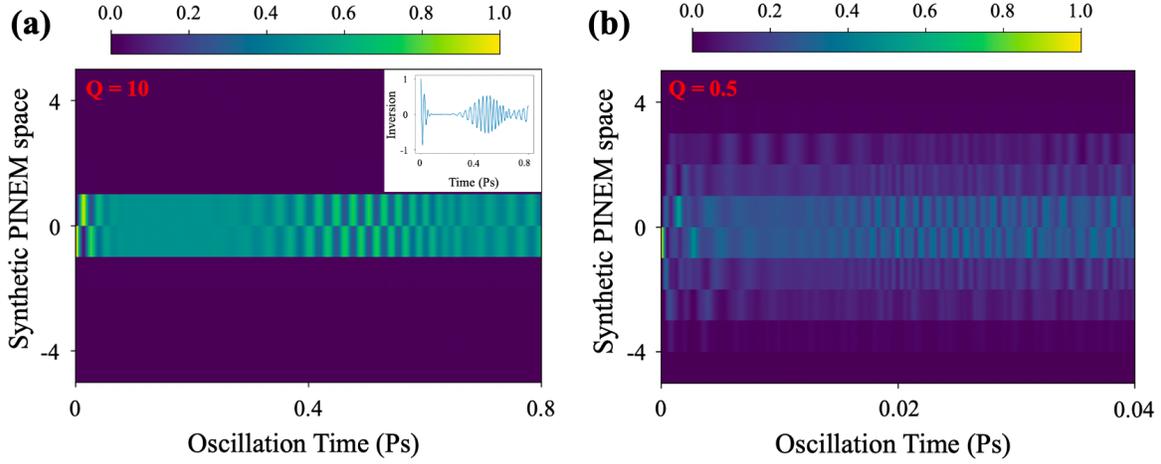

Fig. S3: Exact solution of Eq. S26 in Raman-Nath regime (a) and Bragg regime (b) with a well-defined initial state of the photon (coherent state) and free electron (single sideband $\left|-\frac{1}{2}\right\rangle$). By tuning the intensity of vacuum electric field mode $\tilde{E}_z$, we present the dynamic of the free electron. In the Raman-Nath regime (a), the electron would be trapped in a two-level system and exhibits the collapse and revival effect (see in the insert). In the Bragg regime (b), the electron would be scattered to high-order sidebands.

In the interaction picture, we get

$$H_I^{in}(t) = e^{iH_0 t/\hbar} H_I e^{-iH_0 t/\hbar}$$
$$= \hbar g \left( e^{iv_0 q_z \tau_z t} \tau_+ e^{-iv_0 q_z \tau_z t} \right) \left( e^{i\omega a_q^\dagger a_q t} a_q e^{-i\omega a_q^\dagger a_q t} \right) + h.c.$$
$$= \hbar g \left( \tau_+ e^{iv_0 q_z t} \right) \left( a_q e^{-i\omega t} \right) + h.c.$$

If we define the detuning $\Delta = \omega - v_0 q_z$, then we have

$$H_I^{in}(t) = \hbar \left( g \tau_+ a_q e^{-i\Delta t} + g^* \tau_- a_q^\dagger e^{i\Delta t} \right) \tag{S28}$$

And the time evolution of the combined electron-photon state is determined by

$$i\hbar \frac{\partial}{\partial t} |\psi^{in}\rangle = H_I^{in}(t) |\psi^{in}\rangle \tag{S29}$$

The general input photon state can be written as the superposition of Fock state $|\psi_{ph}\rangle = \sum_n c_n |n\rangle$. Then the combined electron-photon state could be written as

$$|\psi^{in}(t)\rangle = \sum_n c_{\frac{1}{2},n}(t)\left|\frac{1}{2},n\right\rangle + c_{-\frac{1}{2},n}(t)\left|-\frac{1}{2},n\right\rangle \quad (S30)$$

where the normalization is given by $\sum_n \left|c_{\frac{1}{2},n}(t)\right|^2 + \left|c_{-\frac{1}{2},n}(t)\right|^2 = 1$. Then, Eq. (S29) would become

$$\begin{aligned}
i\frac{\partial}{\partial t}|\psi^{in}(t)\rangle &= i\frac{\partial}{\partial t}\sum_n c_{\frac{1}{2},n}(t)\left|\frac{1}{2},n\right\rangle + c_{-\frac{1}{2},n}(t)\left|-\frac{1}{2},n\right\rangle \\
&= \sum_n g^* c_{\frac{1}{2},n}(t)\tau_- a_q^\dagger e^{i\Delta t}\left|\frac{1}{2},n\right\rangle + g c_{-\frac{1}{2},n}(t)\tau_+ a_q e^{-i\Delta t}\left|-\frac{1}{2},n\right\rangle \\
&= \sum_n g^*\sqrt{n+1}c_{\frac{1}{2},n}(t)e^{i\Delta t}\left|-\frac{1}{2},n+1\right\rangle + g\sqrt{n}c_{-\frac{1}{2},n}(t)e^{-i\Delta t}\left|\frac{1}{2},n-1\right\rangle
\end{aligned}$$

Thus, we have

$$\begin{aligned}
i\dot{c}_{\frac{1}{2},n}(t) &= g\sqrt{n+1}e^{-i\Delta t}c_{-\frac{1}{2},n+1}(t) \\
i\dot{c}_{-\frac{1}{2},n+1}(t) &= g^*\sqrt{n+1}e^{i\Delta t}c_{\frac{1}{2},n}(t)
\end{aligned} \quad (S31)$$

For the resonant case ($\Delta = 0$), if the initial photon state is a Fock state $|\psi_{ph}\rangle = |m\rangle$, and the initial electron state is $\left|\frac{1}{2}\right\rangle$. Then we have

$$c_{\frac{1}{2},n}(0) = \delta_{nm}, \quad c_{-\frac{1}{2},n}(0) = 0 \quad (S32)$$

and the corresponding solution is

$$\begin{aligned}
c_{\frac{1}{2},m}(t) &= \cos(\sqrt{m+1}|g|t) \\
c_{-\frac{1}{2},m}(t) &= -i\sin(\sqrt{m}|g|t)
\end{aligned} \quad (S33)$$

Thus, the wavefunction in the Schrodinger picture

$$|\psi(t)\rangle = e^{-iH_0 t/\hbar}|\psi^{in}(t)\rangle$$
$$= e^{-i\left(\frac{v_0 q_z}{2}+n\omega\right)t}\cos(\sqrt{n+1}|g|t)\left|\frac{1}{2},n\right\rangle - ie^{-i\left(n\omega-\frac{v_0 q_z}{2}\right)t}\sin(\sqrt{n}|g|t)\left|-\frac{1}{2},n\right\rangle$$

If the initial photon state is a coherent state $|\psi_{ph}\rangle = |\alpha\rangle = e^{-\frac{|\alpha|^2}{2}} \sum_n \frac{\alpha^n}{\sqrt{n!}} |n\rangle$, and the initial electron state is $\left|\frac{1}{2}\right\rangle$, then

$$c_{\frac{1}{2},n}(0) = c_{\frac{1}{2}}(0) c_n(0) = e^{-\frac{|\alpha|^2}{2}} \frac{\alpha^n}{\sqrt{n!}}$$
$$c_{-\frac{1}{2},n}(0) = 0$$
(S34)

Then, the solution is

$$|\psi^{in}(t)\rangle = e^{-\frac{|\alpha|^2}{2}} \sum_n \frac{\alpha^n}{\sqrt{n!}} \left[ \cos(\sqrt{n+1}|g|t) \left|\frac{1}{2}, n\right\rangle - i \sin(\sqrt{n}|g|t) \left|-\frac{1}{2}, n\right\rangle \right]$$
(S35)

So that, for a coherent state of light input, the population $P_{\frac{1}{2}}(t)$ to detect the electron at $\left|\frac{1}{2}\right\rangle$ is given by

$$P_{\frac{1}{2}}(t) = e^{-|\alpha|^2} \sum_n \frac{\alpha^{2n}}{n!} \cos^2(g\sqrt{n+1}\, t).$$
(S36)

Generally, we can define a general initial photon state $|\psi_{ph}\rangle = \sum_n c_n |n\rangle$, followed by a photon statistics $P_n = |c_n|^2$, with $n$ being the Fock state index. Therefore, we obtain the quantum Rabi dynamics of the combined electron-photon state

$$|\Psi(t)\rangle = \sum_n e^{-i\left(n+\frac{1}{2}\right)\omega t} c_{n+1} \left( \cos(g\sqrt{n+1}\, t) \left|\frac{1}{2}, n\right\rangle - \sin(g\sqrt{n+1}\, t) \left|-\frac{1}{2}, n+1\right\rangle \right),$$
(S37)

in which the population $P_{\frac{1}{2}}(t)$ to detect the electron at $\left|\frac{1}{2}\right\rangle$ is given by

$$P_{\frac{1}{2}}(t) = \sum_n |c_{n+1}|^2 \cos^2(g\sqrt{n+1}\, t).$$
(S38)

which is the main result of the quantum Rabi oscillation of a two-level electron, as presented in Eq. 7 in the main text. Fig. S4 depicts the typical inversion ($I(t) = P_{\frac{1}{2}}(t) - P_{-\frac{1}{2}}(t)$) for thermal, coherent, and squeezed coherent states, respectively.

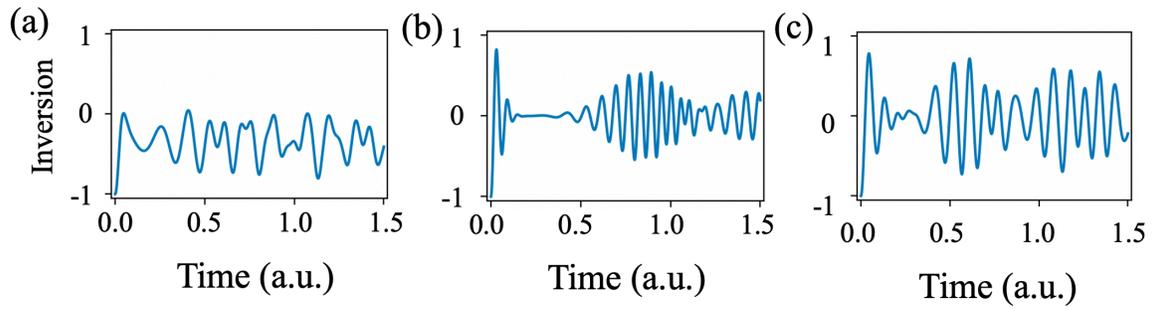

Fig. S4: The time evolution of the inversion $I(t)$ for the thermal, coherent, and squeezed coherent state of light. The Fourier transforms of the inversion are given in Fig. 3 in the main text.